\documentclass[preprint]{rsl}
\title{MeerKAT Reveals Evidence of a Radio Megahalo at GHz Frequency}
\author{Swarna Chatterjee, and Kenda Knowles}

\begin{document}

\maketitle

%
%

\begin{abstract}
Radio megahalos are recently discovered large-scale diffuse synchrotron sources identified through LOFAR observations. We present MeerKAT L-band observations of the massive galaxy cluster RXC~J0528.9$-$3927, that reveals faint emission surrounding its central radio halo. At 1.28 GHz, the known $\sim1.14$ Mpc halo is embedded within previously unreported low-surface-brightness emission extending to $\sim2$ Mpc. The surface-brightness profile of the emission shows a distinct flattening beyond $\sim0.55R_{500}$, suggesting that the outer emission forms an additional component.  These properties make the cluster a candidate megahalo system and potentially the first detected at GHz frequencies, although deeper multi-frequency observations are required for confirmation.

\end{abstract}

\section{Introduction}

\begin{figure*}[htbp]
  \centering
  \includegraphics[width=67mm]{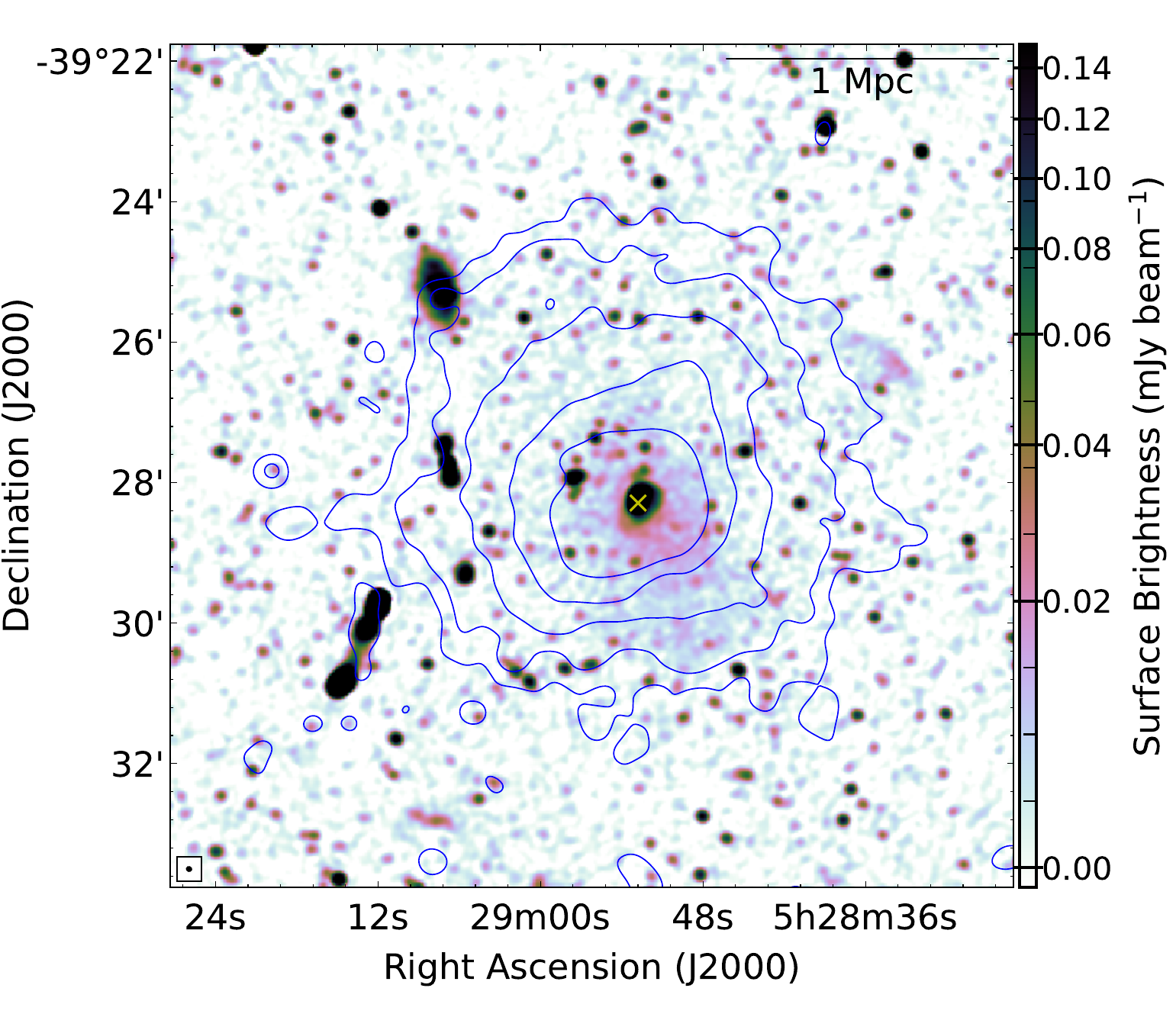}
  \includegraphics[width=67mm]
  {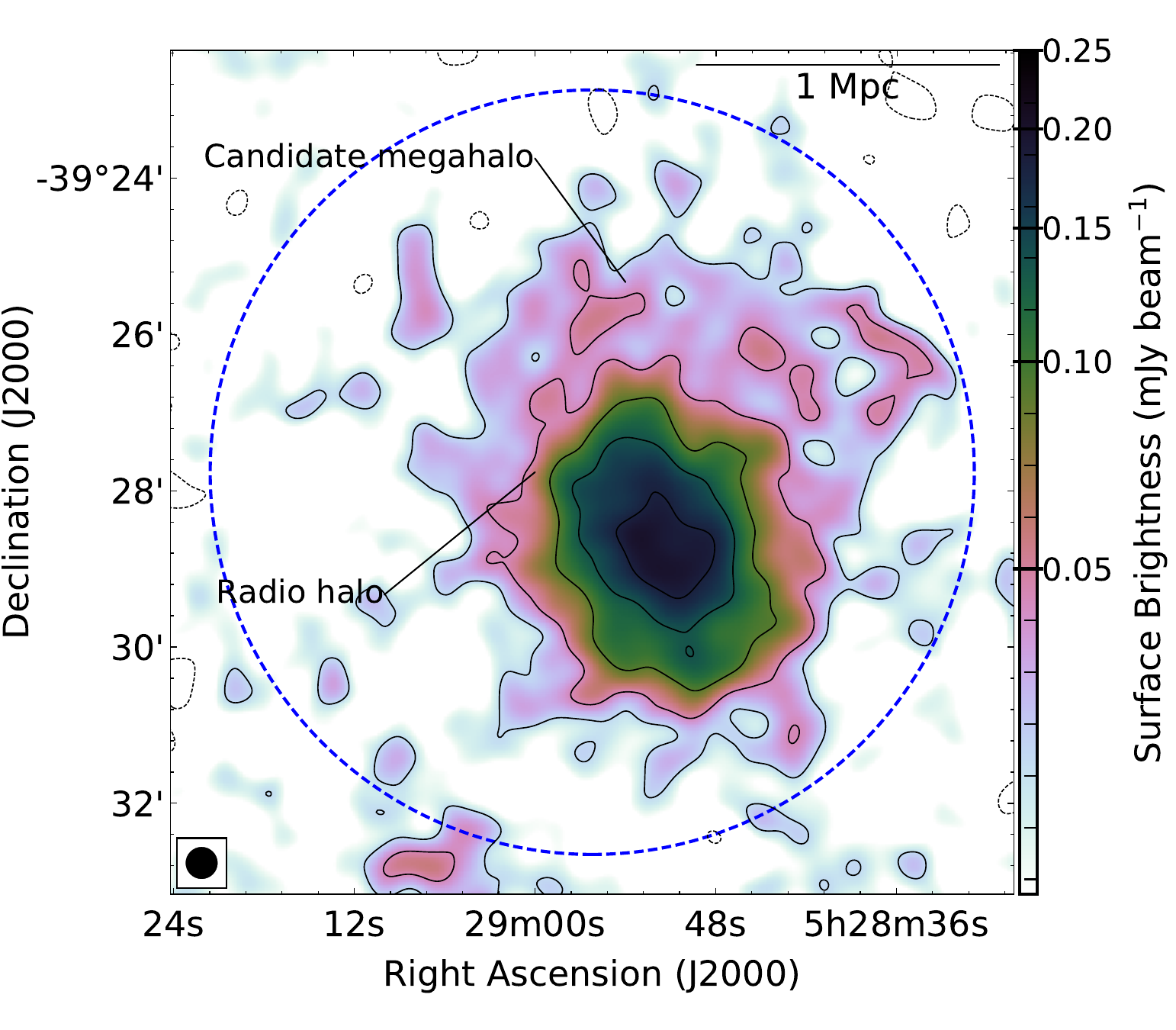}
  \caption{\textit{Left:} Full-resolution ($7.2^{\prime\prime}\times6.7^{\prime\prime}$; PA $=72.0^\circ$) MeerKAT 1.28-GHz image of J0528. The blue contours show the smoothed XMM--Newton surface brightness increasing by factors of two; the yellow 'x' marks the X-ray peak reported by \cite{Foex2017}. \textit{Right:} Compact-source-subtracted MeerKAT image at $26^{\prime\prime}$ resolution overlaid with 1.28 GHz contours placed at $6\times ( \frac{1}{2} , 1, 2, 3, 4, 5) \times 7.5 \mu$ Jy/beam. The dashed blue circle denotes $R_{500}$.}
  \label{fig:asc}
\end{figure*}

In the hierarchical structure-formation scenario, galaxy clusters are the most massive gravitationally bound systems in the Universe. These environments host large-scale diffuse radio emission that is not associated with cluster galaxies. This steep-spectrum synchrotron emission ($\alpha<-1$; $S_{\nu}\propto\nu^{\alpha}$) traces relativistic electrons and magnetic fields in the intracluster medium (ICM), providing insights into large-scale particle acceleration and magnetic-field evolution \cite{Brunetti2014}. Advances in radio interferometry have significantly improved our understanding of diffuse cluster radio emission, leading to the discovery of many fascinating structures, such as radio halos. Radio halos are extended structures found in the central region of galaxy clusters and broadly coincident with the X-ray-emitting region \cite{vanWeeren2019}. Their favoured formation scenario involves the turbulent re-acceleration of pre-existing relativistic electrons during cluster mergers \cite{Brunetti2014}. Recent large-area MeerKAT and LOFAR surveys have revealed further complex radio structures associated with the ICM \cite{Knowles2022,Botteon2022}. One recently discovered class of diffuse radio emission is the radio megahalo. Megahalos extend far beyond classical radio halos and can reach radii comparable to $R_{500}$, the radius within which the mean density is 500 times the critical density of the Universe at the cluster redshift. They occupy volumes up to $\sim30$ times larger and have emissivities approximately 20 times lower than typical halos \cite{Cuciti2022nat}. Their existence suggests that turbulence-driven particle acceleration and cluster magnetic fields extend into the outer ICM, implying that non-thermal components may be widespread on cluster scales. To date, only five megahalos have been reported. Four were identified with LOFAR in massive clusters with $M_{500}>5.58\times10^{14}M_{\odot}$, where $M_{500}$ is the mass enclosed within $R_{500}$, at intermediate redshifts ($0.17\leq z\leq0.28$) \cite{Cuciti2022nat}. The fifth one was discovered through uGMRT observations at 400 and 650 MHz \cite{Salunkhe2025}. Additional targeted observations are essential both to confirm new candidates and to constrain the physical origin of this rare class of diffuse radio sources.

The galaxy cluster \textbf{RXC~J0528.9$-$3927} (hereafter J0528) is a massive system with $M_{500}=(7.41\pm0.37)\times10^{14}\,M_{\odot}$ at $z=0.284$ \cite{Planck2016PSZ2}. It is well studied at optical and X-ray wavelengths and exhibits a bright central X-ray peak, which may indicate a cool core characterised by dense, relatively cool gas with a short radiative cooling time. The central brightest cluster galaxy (BCG) and dynamical analysis suggest that the cluster may be relatively relaxed \cite{Foex2017}. However, the offset of X-ray peak from centre of large scale X-ray emission, two northern galaxy overdensities, and a second bright galaxy $\sim200$ kpc north of the central BCG indicate possible north--south accretion or merger activity, although no major substructure is detected within the virial radius \cite{Foex2017}. Based on its X-ray morphology, \cite{Lovisari2017} classified J0528 having a mixed dynamical state, with characteristics of both relaxed and disturbed systems. In addition, \cite{Botteon2018} reported a cold front in the western region of the cluster, a feature commonly produced by merger-induced gas motions or sloshing. A $\sim1$ Mpc halo has previously been reported in the cluster \cite{Knowles2022,Kolokythas2025}.

Here, we report the first evidence for a potential megahalo detected at high frequency in J0528 using MeerKAT L-band observations. We adopt a flat $\Lambda$CDM cosmology with $\Omega_{\rm m}=0.3$, $\Omega_{\Lambda}=0.7$, and $H_{0}=70$ km s$^{-1}$ Mpc$^{-1}$. At the cluster redshift, $1^{\prime\prime}=4.287$ kpc.

\section{Radio Observations}
We used the MeerKAT L-band (1.28 GHz) archival data for J0528 with 12.9 hours on-source time. The observations were carried out in three epochs with on-source time of 7.5 hours, 1.7 hours, and 3.7 hours, respectively. The data reduction was done with the CARACAL \cite{Jozsa2020} pipeline using standard methods of calibration, flagging and self-calibration. The full resolution map of the cluster (Figure~1; Left) shows diffuse radio emission surrounding the central radio galaxy. To isolate the diffuse emission, we produced a high-resolution model of the compact sources using a uv-range $>$ 1 k$\lambda$ and subtracted the model from the visibility data. The compact source subtracted final diffuse emission map was produced with a $26^{\prime\prime}$ restoring beam (rms $\sim7.5~\mu$Jy~beam$^{-1}$ using WSClean v.3.1.0 \cite{Offringa2014, Offringa2017} (Figure~1; Right).

\section{Results}

At 1.28 GHz, the connected $6\sigma_{\rm rms}$ contour encloses diffuse emission with a largest linear size of $\sim1.14$ Mpc, consistent with the previously reported halo (Figure~1, Right). The map also reveals previously unreported low-surface-brightness emission traced by the connected $3\sigma_{\rm rms}$ contour, extending the total diffuse emission to $\sim1.96$ Mpc. Radio-halo surface-brightness profiles are often described by an exponential model, whereas megahalos show a departure from this behaviour \cite{Cuciti2022nat}. To test whether the outer emission is distinct from the central halo, we followed the fitting procedure of \cite{Cuciti2022nat}, in which a model image with the same dimensions and pixel scale as the observed map was generated and convolved with a Gaussian corresponding to the synthesized beam. For J0528, we adopted the elliptical model used by \cite{Salunkhe2025},
\begin{equation}
I(x,y)=I_{0}\exp\left[-\sqrt{\frac{x^{2}}{r_{1}^{2}}+\frac{y^{2}}{r_{2}^{2}}}\right],
\end{equation}
where $I_{0}$ is the central surface brightness, and $r_{1}$ and $r_{2}$ are the e-folding radii along the semi-major and semi-minor axes, respectively.

The mean surface brightness was measured in elliptical annuli centred on the halo peak and aligned with its morphology. Each annulus had a semi-major-axis width of $13^{\prime\prime}$. Following \cite{Salunkhe2025}, only pixels above $1.5\sigma_{\rm rms}$, and only annuli with a mean surface brightness above $3\sigma_{\rm rms}$ were considered. An elongated feature morphologically consistent with a peripheral radio relic is also detected west--northwest of the cluster and appears connected to the halo in the low-resolution image. However, its nature cannot be established without spectral, polarization, and complementary X-ray constraints. We therefore masked it, together with residual from unrelated sources, when deriving the surface-brightness profile.

Because the outer emission is prominent towards the north, we derived both complete elliptical and northern semi-annular profiles. The annuli, northern region, and source masks are shown in the inset of Figure~\ref{fig:asc}. The beam-convolved single-exponential model assumes a smooth elliptical brightness distribution, whereas the observed halo exhibits small-scale structure. It therefore describes the overall radial decline of the inner halo but does not reproduce every individual measurement. Beyond $r\simeq162.5^{\prime\prime}$ ($\sim697$ kpc), the surface-brightness profile systematically departs from the model. This transition occurs at $\sim0.55R_{500}$, where $R_{500}=1.26$ Mpc, comparable to the radial departures reported for other megahalos \cite{Cuciti2022nat,Salunkhe2025}. Its large extent, low surface brightness, and systematic departure from the inner-halo model therefore motivate its classification as a candidate megahalo.
\begin{figure}[htbp]
  \centering
  \includegraphics[width=0.85\columnwidth]{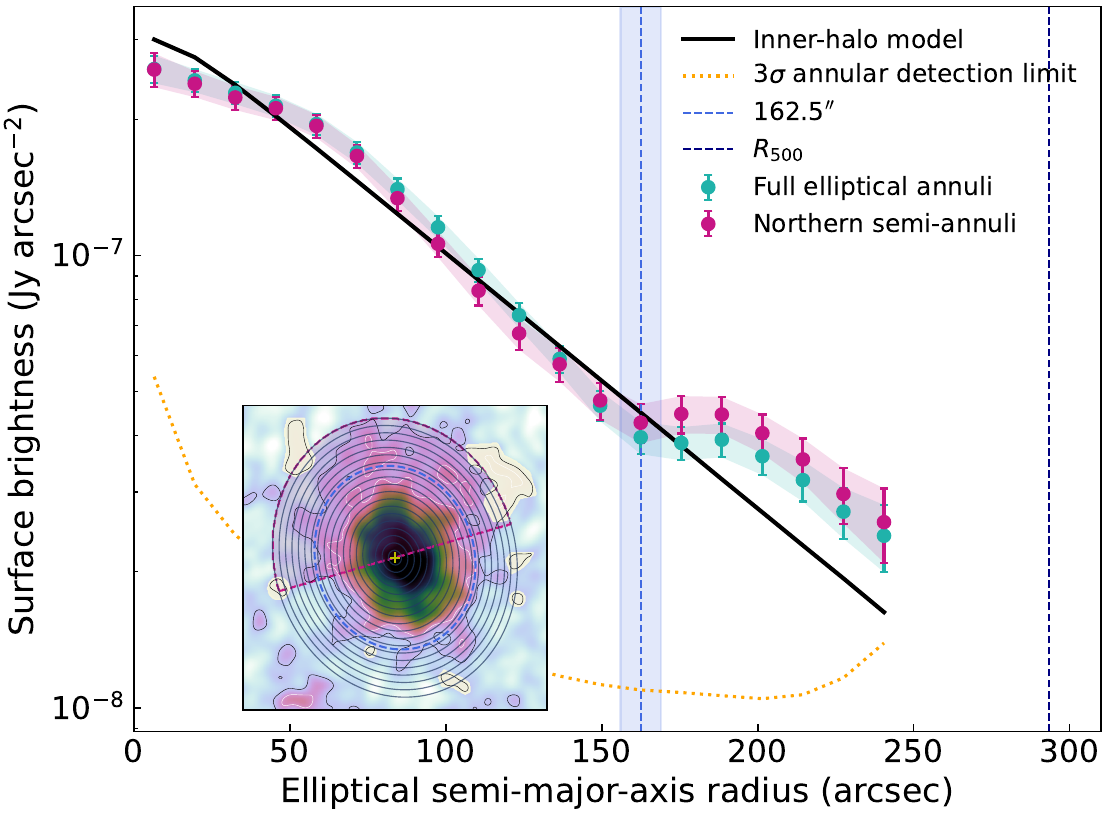}
  \caption{Mean surface-brightness profiles of J0528 using complete elliptical annuli and northern semi-annuli. The shaded regions represent the $1\sigma$ uncertainties, and the solid black line shows the beam-convolved exponential model of the central halo. The yellow dotted curve denotes the $3\sigma$ annular detection limit. The blue dashed lines with shaded region marks the transition radius and the navy dashed line marks the $R_{500}$. The inset shows the annular geometry, selected regions and source masks.
}
  \label{fig:asc}
\end{figure}
As a caveat, we note that low-surface-brightness profiles can be affected by residual compact, embedded, or filamentary sources, potentially producing artificial large-scale components \cite{Rajpurohit2025}. We mitigated this effect by subtracting compact-source models from the visibility data and masking residual extended-source contamination before deriving the profiles.

The central halo enclosed by the connected $6\sigma_{\rm rms}$ contour has an integrated flux density of $S_{\rm halo}=6.53\pm0.33$ mJy. After excluding this region, the directly detected candidate-megahalo emission within the connected $3\sigma_{\rm rms}$ contour has $S_{\rm ext}=2.37\pm0.13$ mJy and a mean surface brightness of $34.08,\mu{\rm Jy,beam^{-1}}$. Assuming that the candidate megahalo extends through the region occupied in projection by the central halo, with the same mean surface brightness measured outside it, we estimate its total flux density of the candidate megahalo to be $S_{\rm mega}=4.48\pm0.25$ mJy.

Assuming a typical radio-halo spectral index of $\alpha=-1.3$ ($S_\nu\propto\nu^\alpha$), we estimated the 1.4-GHz radio power using
\begin{equation}
P_{1.4}=4\pi D_{\rm L}^{2}(1+z)^{-(1+\alpha)}
S_{1.28}\left(\frac{1.4}{1.28}\right)^\alpha ,
\end{equation}
where $D_{\rm L}$ is the luminosity distance and $S_{1.28}$ is the observed flux density at 1.28~GHz. We obtain $P_{\rm halo,1.4}=(1.59\pm0.08)\times10^{24}$ W Hz$^{-1}$ and $P_{\rm mega,1.4}=(1.09\pm0.06)\times10^{24}$ W Hz$^{-1}$.

Megahalos are reported to have steeper spectra ($\alpha\lesssim-1.6$), although this is based on only two systems observed at low frequencies \cite{Cuciti2022nat}. Thus, adopting $\alpha=-1.3$ provides a conservative power estimate. We further computed the volume-averaged emissivity assuming an ellipsoid with image-derived semi-axes $a$, $b$, and $c=b$.

\begin{equation}
j_{1.4}=\frac{P_{1.4}}{\frac{4}{3}\pi abc},
\end{equation}
We obtain $V_{\rm halo}=5.78\times10^{8}$ kpc$^{3}$ and $V_{\rm mega}=2.21\times10^{9}$ kpc$^{3}$, yielding $V_{\rm mega}/V_{\rm halo}=3.83$. The corresponding emissivities are $j_{\rm halo,1.4}=9.4\times10^{-43}$ and $j_{\rm mega,1.4}=1.68\times10^{-43}$ erg s$^{-1}$ Hz$^{-1}$ cm$^{-3}$, making the candidate megahalo $\sim5.6$ times less emissive than the central halo.

\section{Discussion}

The diffuse emission in J0528 extends to $\sim2$ Mpc and shows a surface-brightness flattening beyond the central halo. This indicates the presence of a second low surface-brightness component which is a key observational signature reported for radio megahalos \cite{Cuciti2022nat,Salunkhe2025}. The origin of megahalos remains uncertain. In the framework proposed by \cite{Cuciti2022nat}, major mergers generate strong turbulence in the central cluster region and are generally associated with classical radio halos. By contrast, the continuous accretion of matter may sustain a weak turbulence in the cluster outskirts, which could contribute to the re-acceleration of relativistic electrons and potentially generate megahalo emission. J0528 has been classified as dynamically mixed and hosts a cold front \cite{Foex2017, Botteon2018}, supporting the scenario of an off-axis or minor merger capable of generating turbulence without fully disrupting the core. This is consistent with recent results suggesting a connection between extended diffuse radio emission and sloshing/cold-front features in cool-core clusters \cite{Biava2024}. The presence of a candidate megahalo in such a system therefore supports the idea that cluster-scale non-thermal components may persist even in environments that do not show the most extreme merger signatures.

We stress that the current detection remains tentative owing to the low surface brightness of the extended component and the lack of spectral constraints. Its classification as a candidate megahalo relies primarily on its large extent, low emissivity, and the change in slope of the surface-brightness profile. Previously reported megahalos show steeper spectra than their central radio halos \cite{Cuciti2022nat}; therefore, if physically distinct, the outer component in J0528 would be expected to exhibit spectral steepening near the surface-brightness transition. Deeper multi-frequency observations with matched angular resolution and $uv$ coverage are required to test this prediction and confirm the nature of the emission. Despite these limitations, the present detection is significant for several reasons. First, it represents a unique example of megahalo-like diffuse emission detected at 1.28~GHz, a relatively high frequency where steep-spectrum emission is typically difficult to recover. This demonstrates the capability of deep MeerKAT observations to probe the low surface-brightness non-thermal component of the ICM even at GHz frequencies. Second, the presence of such extended emission in J0528, a system previously classified as dynamically mixed or mildly disturbed \cite{Foex2017, Lovisari2017}, supports the idea that large-scale turbulence and magnetised plasma may persist even in clusters lacking strong merger signatures. If confirmed, this would provide additional evidence that megahalos may not be restricted to the most extreme merger events, but could also arise in clusters undergoing complex or minor merger activity.

\section{Conclusions}
We report faint diffuse radio emission surrounding the central halo in J0528 at 1.28~GHz. Its large extent, low emissivity, and departure from the inner-halo surface-brightness profile suggest this could be a megahalo. Although smaller than the most prominent megahalos reported to date, its detection at GHz frequencies highlights that megahalo-like emission
is not necessarily restricted to very low-frequency observations. The current sample of confirmed megahalos remains extremely small, with confirmed systems so far reported at lower frequencies. Although widefield deep radio surveys are essential for identifying new candidate systems, survey data alone may not always be adequate to robustly recover such faint, large-scale diffuse emission, especially at GHz frequencies. Targeted deeper observations will therefore be crucial for robust confirmation and for establishing the full extent, morphology, and physical nature of this faint large-scale emission. In the near future, the increased sensitivity and survey capability of the Square Kilometre Array (SKA) will bring a transformative improvement, enabling systematic detection and characterisation of faint megahalo emission and offering new insights into particle acceleration and magnetic field evolution in the
intracluster medium.

\section{Acknowledgements}
We thank Rhodes University for supporting this research. We acknowledge funding support from the South African National Research Foundation (NRF). The MeerKAT telescope is operated by the South African Radio Astronomy Observatory, a facility of the NRF, an agency of the Department of Science and Innovation.

\noindent\small
Authors are with Centre for Radio Astronomy Techniques and Technologies, Department of Physics and Electronics, Rhodes University, Artillery Road, Makhanda 6140, South Africa; \\
e-mail: swarna.chatterjee@ru.ac.za.\\
e-mail: k.knowles@ru.ac.za.\\


\begin{thebibliography}{99}

\bibitem{Brunetti2014}
G. Brunetti and T. W. Jones, ``Cosmic Rays in Galaxy Clusters and Their Nonthermal Emission,'' \emph{International Journal of Modern Physics D}, 23, 4, March 2014, 1430007.

\bibitem{vanWeeren2019}
R. J. van Weeren et al., ``Diffuse Radio Emission from Galaxy Clusters,'' \emph{Space Science Reviews}, 215, 1, February 2019, 16.

\bibitem{Knowles2022}
K. Knowles et al., ``The MeerKAT Galaxy Cluster Legacy Survey,'' \emph{Astronomy and Astrophysics}, 657, January 2022, A56.

\bibitem{Botteon2022}
A. Botteon et al., ``The Planck Clusters in the LOFAR Sky. I. LoTSS-DR2: New Detections and Sample Overview,'' \emph{Astronomy and Astrophysics}, 660, April 2022, A78.

\bibitem{Cuciti2022nat}
V. Cuciti et al., ``Galaxy Clusters Enveloped by Vast Volumes of Relativistic Electrons,'' \emph{Nature}, 609, September 2022, pp. 911--914.

\bibitem{Salunkhe2025}
S. Salunkhe, R. Santra, and R. Kale, ``Discovery of a Radio Megahalo in the Cluster PLCKG287.0+32.9 Using the uGMRT,'' \emph{The Astrophysical Journal Letters}, 984, 1, May 2025, L26.

\bibitem{Planck2016PSZ2}
Planck Collaboration, ``Planck 2015 Results. XXVII. The Second Planck Catalogue of Sunyaev--Zeldovich Sources,'' \emph{Astronomy and Astrophysics}, 594, October 2016, A27.

\bibitem{Foex2017}
G. Fo{\"e}x, G. Chon, and H. B{\"o}hringer, ``From the Core to the Outskirts: Structure Analysis of Three Massive Galaxy Clusters,'' \emph{Astronomy and Astrophysics}, 601, May 2017, A145.

\bibitem{Lovisari2017}
L. Lovisari et al., ``X-Ray Morphological Analysis of the Planck ESZ Clusters,'' \emph{The Astrophysical Journal}, 846, 1, September 2017, 51.

\bibitem{Botteon2018}
A. Botteon, F. Gastaldello, and G. Brunetti, ``Shocks and Cold Fronts in Merging and Massive Galaxy Clusters,'' \emph{Monthly Notices of the Royal Astronomical Society}, 476, 4, June 2018, pp. 5591--5620.

\bibitem{Kolokythas2025}
K. Kolokythas et al., ``The MeerKAT Galaxy Cluster Legacy Survey--II. Catalogue of the Diffuse Radio Emission in MeerKAT-GCLS Clusters,'' \emph{Monthly Notices of the Royal Astronomical Society}, 543, 2, October 2025, pp. 1638--1704.

\bibitem{Jozsa2020}
G. I. G. J{\'o}zsa et al., ``CARACal: Containerized Automated Radio Astronomy Calibration Pipeline,'' \emph{Astrophysics Source Code Library}, June 2020, ascl:2006.014.

\bibitem{Offringa2014}
A. R. Offringa et al., ``WSCLEAN: An Implementation of a Fast, Generic Wide-Field Imager for Radio Astronomy,'' \emph{Monthly Notices of the Royal Astronomical Society}, 444, 1, October 2014, pp. 606--619.

\bibitem{Offringa2017}
A. R. Offringa and O. Smirnov, ``An Optimized Algorithm for Multiscale Wideband Deconvolution of Radio Astronomical Images,'' \emph{Monthly Notices of the Royal Astronomical Society}, 471, 1, October 2017, pp. 301--316.

\bibitem{Rajpurohit2025}
K. Rajpurohit et al., ``Radial Profiles of Radio Halos in Massive Galaxy Clusters: Diffuse Giants over 2 Mpc,'' \emph{The Astrophysical Journal}, 992, 1, October 2025, 88.

\bibitem{Biava2024}
N. Biava et al., ``First Evidence of a Connection between Cluster-Scale Diffuse Radio Emission in Cool-Core Galaxy Clusters and Sloshing Features,'' \emph{Astronomy and Astrophysics}, 686, June 2024, A82.

\end{thebibliography}
\end{document}